\documentclass[aps,prb,amsfonts,amsmath,amssymb,superscriptaddress,showpacs,reprint]{revtex4-1}

\usepackage{graphicx}
\usepackage{upgreek}
\usepackage{bm}
\usepackage{xspace}
\usepackage{multirow}
\usepackage{dcolumn}
\usepackage{color}

\newcommand{\Si}[1]{\ensuremath{^{#1}\text{Si}}\xspace}
\newcommand{\natSi}{\ensuremath{^\text{nat}\text{Si}}\xspace}
\newcommand{\Ph}{\ensuremath{{^\text{31}\text{P}}}\xspace}
\newcommand{\As}{\ensuremath{{^\text{75}\text{As}}}\xspace}
\newcommand{\Sblow}{\ensuremath{{^\text{121}\text{Sb}}}\xspace}
\newcommand{\Sbhigh}{\ensuremath{{^\text{123}\text{Sb}}}\xspace}
\newcommand{\Bi}{\ensuremath{{^\text{209}\text{Bi}}}\xspace}

\newcommand{\EM}{\ensuremath{m^*}\xspace}

\newcommand{\DC}{\ensuremath{\epsilon_0}\xspace}

\newcommand{\EBR}{\ensuremath{a_\text{B}^*}\xspace}
\newcommand{\abs}[1]{\ensuremath{\left|#1\right|}\xspace}
\newcommand{\hf}{\ensuremath{A}\xspace} 
 
\newcommand{\amu}{\ensuremath{\text{u}}\xspace}
\newcommand{\gele}{\ensuremath{g_\text{e}}\xspace}
\newcommand{\gnuc}{\ensuremath{g_\text{n}}\xspace}

\newcommand{\muB}{\ensuremath{\mu_\text{B}}\xspace}
\newcommand{\muN}{\ensuremath{\mu_\text{N}}\xspace}
\newcommand{\vect}[1]{\ensuremath{\bm{#1}}\xspace}

\newcommand{\partderivinline}[2]{\ensuremath{\partial{#1}/\partial{#2}}\xspace}

\newcommand{\peaklabel}{\ensuremath{L}\xspace}

\newcommand{\Mbulk}{\ensuremath{M_{\text{bulk}}}\xspace}
\newcommand{\dAdMone}[1][D]{\ensuremath{\alpha_{\text{NN}}^{#1}}\xspace}
\newcommand{\dAdMtwo}[1][D]{\ensuremath{\alpha_{\text{bulk}}^{#1}}\xspace}
\newcommand{\dAdMnet}[1][D]{\ensuremath{\alpha_{\text{net}}^{#1}}\xspace}

\newcommand{\Atot}[1][D]{\ensuremath{A^{#1}}\xspace}

\newcommand{\Mshell}[1]{\ensuremath{M^{(1)}}\xspace}
\newcommand{\Mone}{\ensuremath{M_\text{NN}}\xspace}
\newcommand{\Mtwo}{\ensuremath{M_\text{bulk}}\xspace}

\newcommand{\twentyeight}{A\xspace}
\renewcommand{\natural}{D\xspace}
\newcommand{\twentyeightthirty}{G\xspace}
\newcommand{\twentynine}{H\xspace}
\newcommand{\thirty}{I\xspace}

\begin{document}

\title{Host isotope mass effects on the hyperfine interaction of group-V donors in silicon}

\author{T.~Sekiguchi}
\affiliation{Department of Applied Physics and Physico-Informatics, Keio University, Yokohama 223-8522, Japan}

\author{A.~M.~Tyryshkin}
\affiliation{Department of Electrical Engineering, Princeton University, Princeton, New Jersey 08544, USA}

\author{S.~Tojo}
\affiliation{Department of Applied Physics and Physico-Informatics, Keio University, Yokohama 223-8522, Japan}

\author{E.~Abe}
\altaffiliation[Present address: ]{RIKEN Center for Emergent Matter Science (CEMS), Wako, Saitama, 351-0198, Japan}
\affiliation{Department of Applied Physics and Physico-Informatics, Keio University, Yokohama 223-8522, Japan}

\author{R.~Mori}
\altaffiliation[Present address: ]{Graduate Group in Applied Science and Technology, University of California, Berkeley, California 94720, USA}
\affiliation{Department of Applied Physics and Physico-Informatics, Keio University, Yokohama 223-8522, Japan}

\author{H.~Riemann}
\author{N.~V.~Abrosimov}
\affiliation{Leibniz-Institut f\"{u}r Kristallz\"{u}chtung, D-12489 Berlin, Germany}

\author{P.~Becker}
\affiliation{Physikalisch-Technische Bundesanstalt, D-38116 Braunschweig, Germany}

\author{H.-J.~Pohl}
\affiliation{VITCON Projectconsult GmbH, D-07745 Jena, Germany}

\author{J.~W.~Ager}
\author{E.~E.~Haller}
\affiliation{Lawrence Berkeley National Laboratory, Berkeley, California 94720, USA}

\author{M.~L.~W.~Thewalt}
\affiliation{Department of Physics, Simon Fraser University, Burnaby, British
Columbia V5A 1S6, Canada}

\author{J.~J.~L.~Morton}
\affiliation{London Centre for Nanotechnology, University College London, London WC1H 0AH, UK}

\author{S.~A.~Lyon}
\affiliation{Department of Electrical Engineering, Princeton University, Princeton, New Jersey 08544, USA}

\author{K.~M.~Itoh}
\email{kitoh@appi.keio.ac.jp}
\affiliation{Department of Applied Physics and Physico-Informatics, Keio University, Yokohama 223-8522, Japan}

\date{\today}

\begin{abstract}
The effects of host isotope mass on the hyperfine interaction of group-V donors in silicon are revealed by pulsed electron nuclear double resonance (ENDOR) spectroscopy of isotopically engineered Si single crystals. 
Each of the hyperfine-split \Ph, \As, \Sblow, \Sbhigh, and \Bi ENDOR lines splits further into multiple components, whose relative intensities accurately match the statistical likelihood of the nine possible average Si masses in the four nearest-neighbor sites due to random occupation by the three stable isotopes \Si{28}, \Si{29}, and \Si{30}. 
Further investigation with \Ph donors shows that the resolved ENDOR components shift linearly with the bulk-averaged Si mass. 
\end{abstract}%


\maketitle

With the rapid advancement in studies of phosphorus donor electron spins ($S\!=\!1/2$) and nuclear spins ($I\!=\!1/2$ of \Ph) as potential qubits in silicon-based quantum information processing,\cite{Kane1998_proposal,Zwanenburg2013,Pla2012,Pla2013} the isotope engineering of host silicon has become very important. 
Elimination of the host \Si{29} nuclear magnetic moments by nuclear-spin-free \Si{28} isotopic enrichment led to spectral narrowing of phosphorus donor electron spin resonance (ESR) lines\cite{Abe2004_ESEEM,Abe2010_linewidth,PAM2014_HCT} and the coherence time extension of phosphorus electron spins\cite{Abe2010_linewidth,Tyryshkin2012_T2e} and nuclear spins.\cite{Morton2008_T2n,Steger2012_T2n0,Saeedi2013_T2n+} 
The effects of host silicon isotope (\Si{28}, \Si{29}, \Si{30}) are well known in a variety of impurity electric dipole transitions in silicon.\cite{Cardona2005_mass, Widulle2001_Raman, Yang2006_linewidth, Karaiskaj2013_bindingenergy, Steger2009_group5} 
While the host isotope ``magnetic'' effects, e.g., super-hyperfine interaction of donor electron spins in silicon with surrounding \Si{29} nuclear spins have been investigatd extensively by electron-nuclear double resonance (ENDOR),\cite{Feher1959_ENDOR,Hale1969_SHF} 
previous reports on the host isotope ``mass''  effects on impurity magnetic resonance in silicon are rather limited.\cite{Watkins1975_SnV, Tezuka2010_boron, Stegner2010_boron,Stegner2011_boron-linewidth} 

The present paper reveals the effects of host Si isotope mass composition on the group-V donor hyperfine interaction by ENDOR.\cite{preliminary} 
We show that the Si isotope mass composition modifies the donor ENDOR spectra in two ways: (i) splitting of the ENDOR line up to nine components due to a variation of the Si isotope mass at the four nearest neighbor (NN) lattice sites to each donor, and (ii) frequency shifts of such multi-component ENDOR lines between isotopically engineered silicon due to change in the bulk-averaged mass.  Such mass-induced splittings and shifts in ENDOR frequencies have significant implications for quantum information processing when the donor nuclear spins are employed as qubits.  
If a well-defined resonance frequency is needed across the spin ensemble, one should employ monoisotopic \Si{28} or \Si{30} crystals but not the magnetic \Si{29} to avoid decoherence. 
On the other hand, a mixture of \Si{28} and \Si{30} may provide a frequency-wise addressability to  multiple donor nuclear spins placed in different Si mass surroundings. 

Table \ref{tbl:samples} lists the host isotope composition $\vect{f}$ and the bulk-averaged isotope mass \Mbulk of the phosphorus-doped single-crystal Si samples used in the ENDOR experiments. 
\begin{table}[bp]
\caption{\label{tbl:samples}
Isotopically engineered Si:P single crystals used in this study. 
Sample \natural is naturally abundant silicon (\natSi). 
$f_{m}$ ($m=28,29,30$) is the fractions of the stable isotopes \Si{m} determined by secondary ion mass spectroscopy (SIMS). 
\Mbulk was determined by these $f_m$'s with the individual isotope mass $M_m$ listed in Ref. \onlinecite{Audi2003_nuclearmass}. 
}
\begin{ruledtabular}
\begin{tabular}{crrrc}
Sample &$f_{28}$ (\%) &$f_{29}$ (\%) &$f_{30}$ (\%) &$\Mbulk$ (\amu)\cr
\hline
A\hphantom{$^{a}$}
		  & 99.991
		  &  0.005
		  &  0.004
		  & 27.977\cr
B\hphantom{$^{a}$}
			& 99.920
			&  0.075
			&  0.005
			& 27.978\cr
C\footnotemark[1]
		  & 98.08\hphantom{0}
		  &  1.18\hphantom{0}
		  &  0.74\hphantom{0}
		  & 28.003\cr
D\hphantom{$^{a}$}
		  & 92.23\hphantom{0}
			&  4.67\hphantom{0}
			&  3.10\hphantom{0}
			& 28.086\cr
E\footnotemark[1]
			& 87.19\hphantom{0}
			& 10.28\hphantom{0}
			&  2.53\hphantom{0}
			& 28.130\cr
F\footnotemark[1]
			& 50.27\hphantom{0}
			& 47.87\hphantom{0}
			&  1.86\hphantom{0}
			& 28.492\cr
G\hphantom{$^{a}$}
			& 57.23\hphantom{0}
			&  3.58\hphantom{0}
			& 39.19\hphantom{0}
			& 28.795\cr
H\footnotemark[1]
		  &  0.56\hphantom{0}
		  & 99.23\hphantom{0}
		  &  0.21\hphantom{0} 
		  & 28.973\cr
I\hphantom{$^{a}$}
			&  0.08\hphantom{0}
			&  0.19\hphantom{0}
			& 99.73\hphantom{0}
			& 29.970\cr
\end{tabular}
\footnotetext[1]{Used in Ref. \onlinecite{Abe2010_linewidth}.}
\end{ruledtabular}
\end{table}%
The samples \natural, \twentynine, and \thirty were Czochralski-grown while the rest were float-zone grown,
using the methods described in Ref. \onlinecite{Itoh2003_Cz} and Ref. \onlinecite{Takyu1999_FZ}, respectively. 
The phosphorus donor concentrations were maintained at around $1\!\times\!10^{15}$ $\text{cm}^{-3}$. 
Three other isotopically natural Si crystals (\Si{\text{nat}}) doped with arsenic (\As), antimony (\Sblow and \Sbhigh), and bismuth (\Bi)\cite{Riemann2006_Bi} were studied as well.
Pulsed ENDOR experiments were performed using Bruker Elexsys580 spectrometer at X band (9.7 GHz) equipped with a helium-flow cryostat. 
We used a Davies ENDOR pulse sequence modified with an additional RF pulse (\textit{tidy} pulse) at the end of the sequence to promote a nuclear spin thermalization.\cite{Tyryshkin2006_tidyDavies, Morton2008_tidyDavies} 
For each sample, the lengths of RF pulses were adjusted to be long enough to avoid instrumental broadening of the detected ENDOR lines. 
Temperatures in the range 4.8 -- 8 K were used for \Ph, 4.2 K for \As, 6 K for \Sblow and \Sbhigh, and 15 K for \Bi donors. 
In the case of \Ph donors the ENDOR lineshape was observed to be temperature-invariant below 8 K. 
For temperatures below 5 K (when electron spin $T_1$ relaxation was longer than 10 s), a light emitting diode (LED, 1050 nm) was flashed for 20 ms after each electron-spin echo measurement to accelerate thermalization of electron spins. 
The \textit{tidy} RF pulse in this case was applied in the middle of the LED flash. 
The static magnetic field $B_0$ was applied along the $\left<001\right>$ crystal axis. 
Other crystal orientations ($\left<110\right>$ and $\left<111\right>$) were also examined to confirm no orientation dependence in the ENDOR lineshapes.

All the group-V donors in Si have an electron spin of $S=1/2$ coupled to the non-zero nuclear spin $I$ of the donor nucleus: 
$I=1/2$ for \Ph, $3/2$ for \As, $5/2$ for \Sblow, $7/2$ for \Sbhigh, and $9/2$ for \Bi. 
The spin Hamiltonian with $B_0$ along the $z$ axis is described by $\mathcal{H}=\gele\muB\,B_{0}S_{z}-\gnuc\muN\,B_{0}I_{z}+hA\vect{S}\!\cdot\!\vect{I}$, consisting of the electron and nuclear Zeeman interactions as well as the Fermi contact hyperfine interaction between the electron and nuclear spins with a parameter \hf, which we give in frequency units. 
Solving this spin Hamiltonian to the first order predicts the ENDOR frequencies at $A$ as $\nu=\abs{A m_S-\gnuc\muN\,B_0/h}$ where $m_S$ is the electron spin projection.  While the hyperfine parameter \hf for each donor in \natSi is well known, e.g., $\hf=117.5$ MHz for \Ph,\cite{Feher1959_ENDOR} the present work reveals a significant dependence of \hf on the host Si isotope composition. 
\begin{figure}[ht!] 
\includegraphics[width=0.45\textwidth]{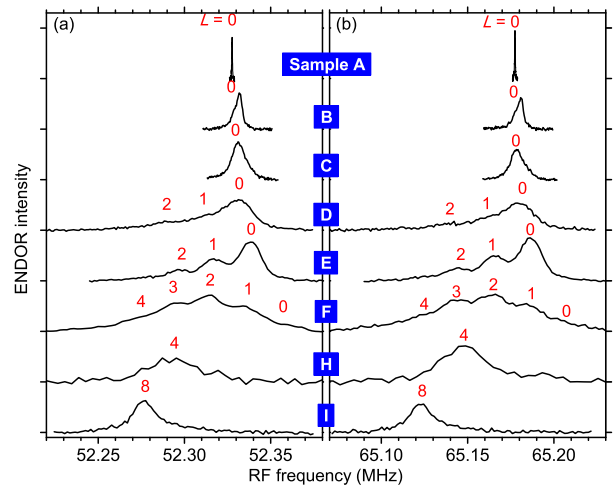}
\caption{\label{fig:OverallSpectra}(color online). 
\Ph ENDOR spectra of the 
(a) $m_S=+1/2$ and (b) $m_S=-1/2$ states 
for eight samples at the high-field (the nuclear spin projection $m_I=-1/2$) ESR line. 
\Mbulk increases from top to bottom. 
A spectrum of \twentyeightthirty is shown independently in Fig. 2 (d) since it is too broad to fit in here and shifted slightly from others due to a small different in the microwave excitation frequency employed in that particular measurement. 
Peaks are labeled with integers $\peaklabel=0$ to $8$ based on the change in the average mass of the four NN Si isotopes, \Mone, as described later in the text. 
}
\end{figure}%

\begin{figure}[ht!] 
\includegraphics[width=0.475\textwidth]{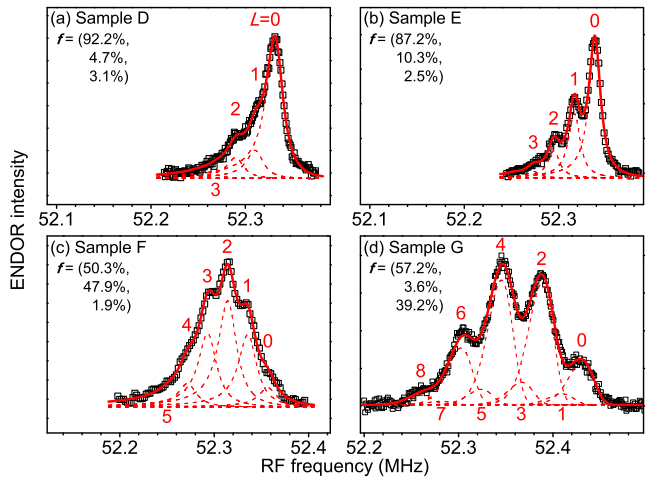}
\caption{\label{fig:peakfitting28+29+30}(color online). 
Fitting of the ENDOR spectra for samples (a) \natural, (b) E, (c) F, and (d) \twentyeightthirty. 
The open squares (black) are experimental data. 
The solid and dashed curves (red) are fitting results of the full spectra and individual components ($L=0$ to $8$), respectively. 
The relative intensities of the components are determined solely by the statistical probability of \Mone. 
The lineshape is optimized as common to all the components: asymmetric Gaussian for (d) and asymemtric Lorentzians for the rest.\cite{Stancik2008_asymmetry} 
}
\end{figure}%
Figures \ref{fig:OverallSpectra} and \ref{fig:peakfitting28+29+30} show a variation of the \Ph ENDOR spectra for the nine isotopically different Si crystals. 
From these spectra we immediately find the following
host isotope effects: 
(i) The ENDOR line of isotopically mixed samples splits into multiple components labeled $L=0, 1, \cdots, 8$. 
(ii) Each component $L$ shifts upwards with increasing \Mbulk. 
The spectra for each sample are identical between (a)  $m_S=+1/2$ and (b) $-1/2$ in Fig. \ref{fig:OverallSpectra}, 
indicating that the observations (i) and (ii) are not due to changes in the electron and nuclear $g$-factors. 
We also exclude the ``magnetic'' host isotope effect due to \Si{29} nuclear spins since, as seen in Fig. \ref{fig:peakfitting28+29+30}(d), sample \twentyeightthirty that has the largest mass-disorder with non-magnetic \Si{28} and \Si{30} with a relatively small amount of magnetic \Si{29} shows the largest degree of splitting spanning $L=0$ to  $8$.
Indeed, as we will show below, the following equation including the mass effects only is found to describe the experimentally observed host isotope effects on the donor dependent hyperfine parameter:
\begin{align}\label{eq:A_Mass}
\Atot = A_{28}^{D} +\dAdMone\,(\Mone-M_{28}) +\dAdMtwo(\Mtwo-M_{28}),
\end{align}%
where $A_{28}^{D}$ is the hyperfine parameter for a specified donor $D$ (\Ph, \As, \Sblow, \Sbhigh, or \Bi) 
in a monoisotopic \Si{28} crystal. 
The second term, which is proportional to the difference of the average mass \Mone of the four NN Si isotopes from the \Si{28} isotope mass $M_{28}$, 
describes the ENDOR line splitting [observation (i)] with an experimentally obtained \dAdMone independent of \Mbulk. 
The third term represents the contribution of the ``bulk'' effect [observation (ii)].  
This is proportional to the mass difference of \Mbulk from $M_{28}$ and describes the ENDOR frequency shift with an experimentally obtained \dAdMtwo independent of \Mone. 

Let us begin our analysis from the NN mass effect. The labeling $\peaklabel$ in Figs. 1 and 2 is given as follows.  
Taking into account that there are four NN sites to each donor and 
that the mass differences between the three stable Si isotopes are equal to $1.00$~\amu, \Mone can take only nine different values between $M_{28}$ and $M_{30}$
in an increment of $0.25$~\amu, 
which correspond to the resolved lines labeled with integers $\peaklabel$ between $0$ and $8$, 
i.e., $\peaklabel =  \left(\Mone-M_{28}\right)/\left(0.25\,\amu\right)$. 
The statistical distribution of the three kinds of Si isotopes in the NN sites is determined by the isotopic fractions $(f_{28},f_{29},f_{30})$ in each sample as given in Table \ref{tbl:samples}. 
Therefore, the relative ENDOR intensity of each $\peaklabel$ in each sample can be estimated as the sum of the distribution probabilities of the NN Si isotopic configurations having the corresponding \Mone. 
Figure \ref{fig:peakfitting28+29+30} shows 
the fitting results based on this NN mass effect 
using a single splitting parameter $\dAdMone[\text{P}]$ for each sample 
with a common optimized lineshape for all the components. 
The calculated distributions of \Mone reproduce the experimentally measured relative intensities very well. 
The line splitting is found to be $\Delta \nu_{\text{EN}} = -21(2)$ kHz per $\Delta L=1$, 
leading to 
$\dAdMone[\text{P}] 
= 4(\Delta\hf/\Delta\peaklabel)
= -170(6)\,\text{kHz}/\amu,
$
or $\dAdMone[\text{P}]/A=-1.45(5)\times10^{-3}\,\amu^{-1}$. 
Note that the binding energies of the S$^+$ and Se$^+$ donors have also negative linear dependences on \Mone.\cite{Pajot2004_S+bindingenergy, Steger2009_group6}

\begin{figure}[ht!]
\includegraphics[width=0.475\textwidth]{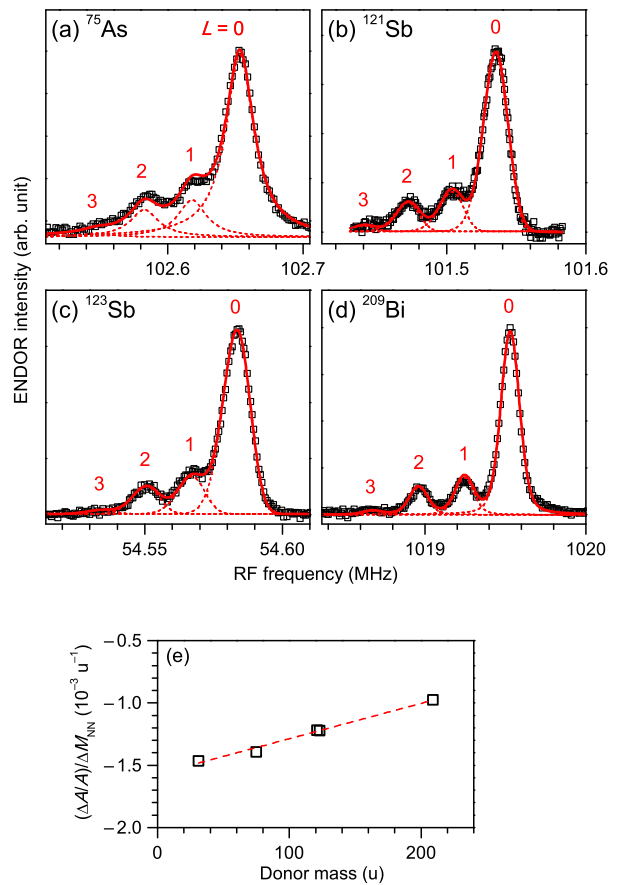}
\caption{\label{fig:otherDonors}(color online). 
The ENDOR spectra of (a) \As, (b) \Sblow, (c) \Sbhigh, and (d) \Bi donors in \Si{\text{nat}}. 
The open squares (black) are experimental data. 
The solid and dashed curves (red) are fitting results as employed in Fig. \ref{fig:peakfitting28+29+30} with an optimized lineshape for all the components: 
asymmetric Lorentzian for (a), and asymmetric Gaussians for the other spectra.  
Each spectrum corresponds to the NMR transition of 
$m_I$ = 
$(-1/2\rightarrow+1/2)$, 
$(-5/2\rightarrow-3/2)$, 
$(-7/2\rightarrow-5/2)$, and
$(-7/2\rightarrow-5/2)$, 
respectively. 
The same labeling scheme $L$ as in Fig. \ref{fig:peakfitting28+29+30} is used to identify the peaks. 
(e) The fractional change of the hyperfine parameter with respect to \Mone (open squares) plotted as a function of the donor isotope mass along with a linear fit (dashed line) with a slope of $3\times10^{-6}\,\amu^{-2}$. 
}
\end{figure}

Similar ENDOR splitting patterns are observed for the other group-V donors in \Si{\text{nat}} as shown in Figs. \ref{fig:otherDonors}(a)--\ref{fig:otherDonors}(d), demonstrating that the NN Si isotope mass effect on \hf is universal. 
A linear dependence of the fractional change in \hf on the donor isotope mass, 
i.e., $(\Delta\Atot/\Atot)/\Delta\Mone$, 
is found as shown in Fig. \ref{fig:otherDonors}(e). 
The value for \Bi 
agrees with the observation by Fourier transform ESR at a magnetic-field clock transition.\cite{Wolfowicz2013_bismuthCT} 

A question remains why it is possible to describe $L=0$ -- $8$ splitting by the average mass \Mone without considering the configuration of the isotopes within the four NN sites. 
For example the $L=4$ peak with $\Mone = 29.0\,\amu$ includes the contributions from the four NN sites occupied by (\Si{29}, \Si{29}, \Si{29}, \Si{29}), (\Si{28}, \Si{28}, \Si{30},  \Si{30}), and (\Si{28}, \Si{29}, \Si{29}, \Si{30}). Such differences in the combination are not resolved for any $L$ within our experimental condition. 
Although it may be the case that specific NN symmetries and/or vibrational modes\cite{Pajot2004_S+bindingenergy} are playing important roles, further investigation is needed to resolve this issue.  

Let us now discuss the bulk-averaged mass effect on \hf, i.e., the third term in Eq. \eqref{eq:A_Mass}. 
\begin{figure}[ht!]
\includegraphics[width=0.45\textwidth]{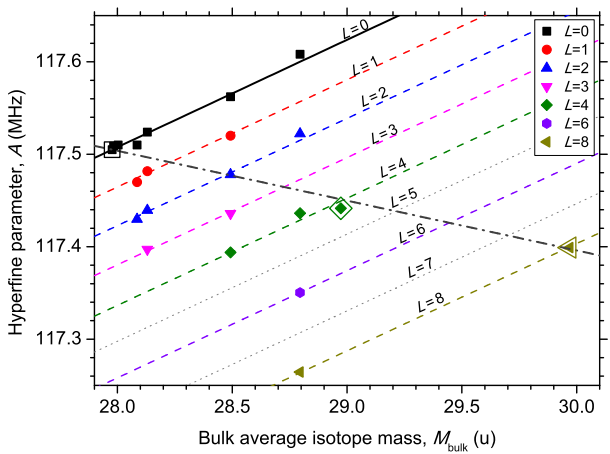}
\caption{\label{fig:EndorFrequencies}(color online). 
The hyperfine parameter \Atot[\text{P}] for \Ph as a function of $\Mbulk$. 
The solid line is a linear fitting of $\peaklabel=0$ yielding the slope $+116(8)\,\text{kHz}/\amu$ and $A_{28}^\text{P}=117.505$ MHz. 
Keeping the same slope, fitting with the vertical offset as a parameter is performed for each of $\peaklabel = 1,2,3,4,6,8$ (dashed lines). The dotted line for $\peaklabel=5$ (7) is drawn as the midway parallel line between $\peaklabel=4$ and 6 (6 and 8). 
The dot-dashed line is a linear fitting [$-54(3)\,\text{kHz}/\amu$] to the three monoisotopic samples (\twentyeight, \twentynine, and \thirty), which are highlighted by enlarged open symbols.
}
\end{figure}%
Figure \ref{fig:EndorFrequencies} plots \hf for each resolved \Ph ENDOR component $\peaklabel$ as a function of \Mbulk. 
Here, \hf for each sample has been determined by summing the ENDOR peak frequencies  in Figs. \ref{fig:OverallSpectra}(a) and \ref{fig:OverallSpectra}(b) for each \peaklabel.
The $L=0$ data are fitted by a linear function of \Mbulk (solid line) with a slope of $\dAdMtwo[\text{P}]=+116(8)$ kHz/\amu. 
The linear lines for other $L$'s are drawn as described in the caption of Fig. 4, 
showing the same slope as $L=0$ with an equidistant vertical offset between the adjacent lines. 

It is rather surprising that the present experimental results can be modeled by the linear shifts due to \Mone and \Mbulk separately. 
While the physical origin of these linearity and separability is  unclear, the net mass effect including both the NN and bulk mass effects can be directly seen by focusing on 
the nearly monoisotopic Si:P samples enriched by \Si{28}, \Si{29}, and \Si{30}.
Data from the three samples are highlighted by enlarged open symbols in Fig. \ref{fig:EndorFrequencies}. 
Only one ENDOR component corresponding to $L=0$, $4$, and $8$ appears in the samples \twentyeight, \twentynine, and \thirty, respectively, since all the four NN sites in these samples are occupied predominantly by \Si{28}, \Si{29}, and \Si{30}, respectively. 
Because $\Mone\approx\Mbulk$, the second and third terms in Eq. \eqref{eq:A_Mass} are merged into 
$\dAdMnet[\text{P}](\Mbulk-M_{28})$, 
whose slope 
$\dAdMnet = -54(3)\,\text{kHz}/\amu$ 
is represented by the dot-dashed line in Fig. \ref{fig:EndorFrequencies}.  
As expected, this value is consistent with the sum of the separately obtained slopes, i.e., $\dAdMone[\text{P}]+\dAdMtwo[\text{P}]=-54(14)$ kHz/\amu. 
In contrast, the net mass shifts in binding energy are positive for the P and S$^+$ donors.\cite{Steger2009_group6,Steger2009_group5} 

Because the Fermi contact hyperfine parameter \hf is proportional to 
the electron density at the nucleus, by using the ground state envelope function $\Phi(r)\propto{\EBR}^{-3/2}\exp\left(-r/\EBR\right)$ with the effective Bohr radius \EBR defined by the static dielectric constant \DC and the electron effective mass \EM,\cite{Yu2005} 
it behaves as 
\begin{align}\label{eq:AvsDC&EM}
	\hf \propto \abs{\Phi(0)}^2 \propto {\EBR}^{-3} \propto {\DC}^{-3}\,{\EM}^{+3}.
\end{align}%
From this relationship we estimate $\partderivinline{\ln\hf}{\Mbulk}$ to be 
$\left(-3\delta\DC/\DC +3\delta\EM/\EM\right)/\delta\Mbulk$, 
where $\delta$ refers to the change in the values of the respective parameters between \Si{28} and \Si{30}. 
Using $\delta\DC/\DC = 6.5\times10^{-4}$ and $\delta\EM/\EM = -1.1\times10^{-4}$ with $\delta\Mbulk = -2.0$~\amu taken from Ref. \onlinecite{Karaiskaj2013_bindingenergy}, we arrive at  $(\partderivinline{\hf}{\Mbulk})/\hf = 1.1\times10^{-3}\,\amu^{-1}$. 
Here employment of $\hf=15.3$ MHz estimated by the effective mass approximation (EMA)\cite{Kohn1955,Ivey1975} leads to
$\partderivinline{\hf}{\Mbulk} = +18$ kHz/\amu, 
which differs from the experimentally obtained slopes $\dAdMnet[\text{P}] = -54$ kHz/\amu and $\dAdMtwo[\text{P}] = +116$ kHz/\amu. 
Employment of the experimental value $\hf=117.5\,\text{MHz}$ leads to  $\partderivinline{\hf}{\Mbulk} = (\partderivinline{\ln\hf}{\Mbulk})A = +130$ kHz/\amu, which interestingly agrees with the experimentally obtained $\dAdMtwo[\text{P}] = +116$ kHz/\amu of the third term in Eq. (1). 
Our preliminary analysis of \Bi ENDOR with an isotopically enriched \Si{28} sample (data not shown) and \natSi sample [Fig. 3(d)] 
leads to a value $\dAdMtwo[\text{Bi}]/\Atot[\text{Bi}] = 1.24\times10^{-3}\,\amu^{-1}$, 
which is comparable to 
the experimental result for \Ph,
$\dAdMtwo[\text{P}]/\Atot[\text{P}] = 0.99\times10^{-3}\,\amu^{-1}$.
Thus, when the NN mass \Mone is fixed, 
the experimental $\dAdMtwo/\Atot$ 
seems independent of the donor species 
and takes a value similar to $(\partial A/\partial M_\text{bulk})/A=1.1\times10^{-3}\,\amu^{-1}$ expected from Eq. (2).
However, we do not have complete theoretical justification to employ the experimentally found $A$ in the above analysis. 
Rigorous evaluation of our experimental findings requires further theoretical research involving advanced methods.\cite{Gerstmann2011_abinitio,Greenman2013_DFT} 

In conclusion, we have revealed the host isotope mass effects on the hyperfine interaction of group-V donors from the variation in the ENDOR spectra of various isotopically engineered Si crystals. 
The relative intensities of the split ENDOR compoents for all the group-V donors are explained by a negative linear dependence of the hyperfine parameter \hf on the average Si isotope mass \Mone at the four nearest-neighbor sites to the donor. 
The donor isotope mass dependence of the fractional change in \hf by \Mone has been determined. 
An identical positive linear shift of all the split components with the bulk-averaged mass \Mbulk has been identified for \Ph. 
The net Si isotope mass effect on \hf observed directly from the isotopically enriched Si:P samples exhibits a negative linear dependence on the Si isotope mass.

\begin{acknowledgments}
This work was supported in part by the Grant-in-Aid for Scientific Research by MEXT, in part by NanoQuine, in part by FIRST, and in part by JSPS Core-to-Core Program. Isotopically enriched Si crystal growth at  LBL was supported by the Director, Office of Science, Office of Basic Energy Sciences, Materials Sciences and Engineering Division of the U.S. Department of Energy under Contract No. DE-AC02-05CH11231.  
\end{acknowledgments}

\end{document}